\documentclass[12pt]{article}
\usepackage{amsmath}

\begin{document}

\title{{\vspace{-.60cm}\textbf{\Large Quantum mechanical probabilities and \\
\vspace{-.1cm}general probabilistic constraints for \\
\vspace{-.25cm}Einstein--Podolsky--Rosen--Bohm experiments}}}
\vspace{1cm}
\author{Jos\'{e} L. Cereceda  \\ 
\textit{C/Alto del Le\'{o}n 8, 4A, 28038 Madrid, Spain}  \\
\small{Electronic address: jl.cereceda@teleline.es}}

\date{24 September, 2000}

\maketitle
\begin{abstract}
Relativistic causality, namely, the impossibility of signaling at superluminal speeds, restricts the kinds of correlations which can occur between different parts of a composite physical system. Here we establish the basic restrictions which relativistic causality imposes on the joint probabilities involved in an experiment of the Einstein-Podolsky-Rosen-Bohm type. Quantum mechanics, on the other hand, places further restrictions beyond those required by general considerations like causality and consistency. We illustrate this fact by considering the sum of correlations involved in the CHSH inequality. Within the general framework of the CHSH inequality, we also consider the nonlocality theorem derived by Hardy, and discuss the constraints that relativistic causality, on the one hand, and quantum mechanics, on the other hand, impose on it. Finally, we derive a simple inequality which can be used to test quantum mechanics against general probabilistic theories.

\vspace{.3cm}
\noindent \textit{Key words:} EPRB-type experiment, joint probability, causal communication constraint, quantum mechanics,
Bell's inequality, Hardy's nonlocality theorem.
\end{abstract}

\section{Introduction}

It is a remarkable feature of elementary, nonrelativistic quantum mechanics that it does not conflict with the theory of special relativity for any practical purpose. An important example of this ``peaceful coexistence'' between quantum mechanics and special relativity is provided by the fact that one cannot exploit the quantum mechanical correlations taking place between spatially separated parts of a composite quantum system to convey classical messages faster than light [1,2], in spite of the fact that such quantum correlations can yield a violation of Bell's inequality [3]. Indeed, the quantum mechanical probabilities behind those correlations are found to satisfy the so-called causal communication constraint [4] (also referred to in the literature   as the condition of ``parameter independence'' [5], ``simple locality'' [6], ``signal locality'' [7], or ``physical locality'' [8]) which, roughly speaking, stipulates that the probability of a particular measurement outcome on any one part of the system should be independent of which sort of measurement was performed on the other parts. This requirement prevents the acausal exchange of classical information between them. Consider an experimental set-up of the Einstein-Podolsky-Rosen-Bohm type [9,10] designed to test the Clauser-Horne-Shimony-Holt (CHSH) version [11,12] of Bell's inequality. The CHSH inequality concerns a statistical ensemble of identically prepared systems, each of which consisting of two parts (call them, say, $A$ and $B$) far away from one another. Let $a_1$, $a_2$, $b_1$, and $b_2$ denote two-valued ($\pm 1$) physical variables, with $a_1$ and $a_2$ referring to measurements on part $A$ of the system by a local observer, and $b_1$ and $b_2$ referring to local measurements on part $B$. For each of the systems in the ensemble a measurement of either $a_1$ or $a_2$ ($b_1$ or $b_2$) is performed on part $A$ ($B$). A complete experimental run on the ensemble of systems will yield the set of numerical values $p(a_j =m, b_k =n)$, where $j,k = 1 \text{ or } 2$, and $m,n=\pm 1$, with $p(a_j =m, b_k =n)$ being the probability of getting the outcomes $a_j=m$ and $b_k=n$ in a joint measurement of the variables $a_j$ and $b_k$. Each of these probabilities fulfills the property
\begin{equation}
0 \leq  p(a_j =m, b_k =n) \leq 1 .
\end{equation}
Furthermore, the various observable probabilities are assumed to satisfy the normalization condition
\begin{equation}
\sum_{m,n =\pm 1}  p(a_j =m, b_k =n) =1 ,
\end{equation}
for any $j,k = 1 \text{ or } 2$. The causal communication constraint, on the other hand, requires that [4]
\begin{align}
&p(a_j =m) = \sum_{n=\pm 1} p(a_j =m, b_1 =n) = \sum_{n=\pm 1} p(a_j =m, b_2 =n), \tag{3a} \\
&p(b_k =n) = \sum_{m=\pm 1} p(a_1 =m, b_k =n) = \sum_{m=\pm 1} p(a_2 =m, b_k =n). \tag{3b}
\end{align}
Condition (3a) states that the probability of obtaining $a_j =m$ is independent of which measurement ($b_1$ or $b_2$) is performed on part $B$. Similarly, condition (3b) states that the probability for $b_k =n$ is independent of which measurement ($a_1$ or $a_2$) is performed on part $A$.

Let us define the correlation function $c(a_j ,b_k)$ between the variables $a_j$ and $b_k$ to  be the expectation value of the product $a_j b_k$. In terms of the measurable probabilities $p(a_j =m, b_k =n)$, this quantity can be expressed as
\begin{align}
c(a_j ,b_k) = \;\, &p(a_j =1,b_k=1) + p(a_j=-1,b_k=-1)  \nonumber  \\
&-p(a_j=1,b_k=-1) - p(a_j=-1,b_k=1)\,.   \tag{4}
\setcounter{equation}{4}
\end{align}
The CHSH inequality [11,12]
\begin{equation}
-2 \leq c(a_1,b_1) + c(a_1,b_2) + c(a_2,b_1) - c(a_2,b_2) \leq 2 \, ,
\end{equation}
holds in any theory of local hidden variables, and restricts the maximum absolute value of the sum of correlations, $\Delta \equiv c(a_1,b_1) + c(a_1,b_2) + c(a_2,b_1) - c(a_2,b_2)$, to 2. We can write the CHSH inequality in a compact notation as $\left| \Delta_{\text{LHV}} \right| \leq 2$. On the other hand, it is well known since the work of Cirel'son [13] that the quantum prediction of the CHSH sum of correlations is bounded in absolute value by $2\sqrt2$, that is, $\left| \Delta_{\text{QM}} \right| \leq 2\sqrt2$. This latter bound, however, still lies well below the maximum absolute theoretical value, $4$, allowed by a general probabilistic theory, $\left| \Delta_{\text{GP}} \right| \leq 4$. This absolute probabilistic limit is attained whenever $c(a_1,b_1)= c(a_1,b_2)= c(a_2,b_1)= -c(a_2,b_2)= \pm 1$. Popescu and Rohrlich [14] addressed the question of whether relativistic causality restricts the maximum quantum prediction of the CHSH sum of correlations to $2\sqrt2$ instead of 4. As Popescu and Rohrlich put it [14], ``Rather than ask why quantum correlations violate the CHSH inequality, we might ask why they do not violate it \textit{more}.'' They found that relativistic causality does not by itself constrain the maximum CHSH sum of quantum correlations to $2\sqrt2$. Indeed, they gave a set of probabilities which satisfies the causal communication constraint and which provides the maximum level of violation, \nolinebreak 4.

In this paper we will work out the basic relationships which develop between the joint probabilities involved in the CHSH inequality when these are required to satisfy the normalization condition and the causal communication constraint. After this is done, we will focus on the particular case in which \textit{three\/} specific probabilities are equal to zero. This leads us directly to a consideration of Hardy's nonlocality theorem [15,16], and enables us to deduce in a very general and economical way the constraints that probability theory and causality impose on it. We shall see that quantum mechanics imposes further restrictions of its own beyond those required by the causal communication constraint. Furthermore, following a suggestion made by Kwiat and Hardy at the end of their recent paper in Ref.\ 17, we will explore the middle ground between the limits imposed by quantum mechanics and relativistic causality within the context of Hardy's theorem. We note that, interestingly, all this is done within the general framework of the CHSH inequality. This allows a unified treatment of both the CHSH inequality and Hardy's nonlocality theorem, and avoids to deal with two different types of inequalities (for example the CHSH inequality and the Clauser-Horne (CH) inequality [12,18], had Hardy's theorem been cast in this latter form of inequality). Moreover, this unification is convenient because, as shown by Mermin [8], the CHSH and CH inequalities need not be equivalent if the causal communication constraint, Eqs.\ (3), does not hold.

\section{General constraints on the joint probabilities in the CHSH inequality}

The CHSH inequality (5) involves sixteen joint probabilities $p(a_j =m, b_k =n)$, although, as we shall presently see, the constraints in Eqs.\ (2) and (3) reduce the number of independent probabilities to eight. In order to abbreviate the notation, from now on the various probabilities $p(a_j =m, b_k =n)$ will be referred to by the respective shorthands $p1, p2,\ldots,p16$, according to the following convention
\begin{align}
p1 &\equiv p(a_1=1, b_1=1),  & p2 &\equiv p(a_1=1, b_1=-1),  \notag  \\
p3 &\equiv p(a_1=-1, b_1=1), & p4 &\equiv p(a_1=-1, b_1=-1), \notag  \\
p5 &\equiv p(a_1=1, b_2=1),  & p6 &\equiv p(a_1=1, b_2=-1),  \notag   \\
p7 &\equiv p(a_1=-1, b_2=1), & p8 &\equiv p(a_1=-1, b_2=-1),    \\
p9 &\equiv p(a_2=1, b_1=1),  & p10 &\equiv p(a_2=1, b_1=-1),  \notag  \\
p11 &\equiv p(a_2=-1, b_1=1),& p12 &\equiv p(a_2=-1, b_1=-1), \notag  \\
p13 &\equiv p(a_2=1, b_2=1), & p14 &\equiv p(a_2=1, b_2=-1),  \notag  \\
p15 &\equiv p(a_2=-1, b_2=1),& p16 &\equiv p(a_2=-1, b_2=-1). \notag   
\end{align}
Now we can write down explicitly the constraints imposed by the requirements of normalization (cf.\ Eq.\ (2)) and causality (cf.\ Eqs.\ (3a)-(3b)) as follows
\begin{align}
& p1+p2+p3+p4 =1,   \notag  \\
& p5+p6+p7+p8 =1,    \notag  \\
& p9+p10+p11+p12 =1,  \notag  \\
& p13+p14+p15+p16 =1,  \notag \\
& p1+p2-p5-p6 =0,     \notag \\
& p3+p4-p7-p8 =0,      \\
& p9+p10-p13-p14 =0,   \notag \\
& p11+p12-p15-p16 =0,   \notag  \\
& p1+p3-p9-p11 =0,   \notag \\
& p2+p4-p10-p12 =0,   \notag  \\
& p5+p7-p13-p15 =0,   \notag \\
& p6+p8-p14-p16 =0.  \notag
\end{align}
Relations (7) constitute a system of 12 linear equations with 16 unknowns. It can be shown that the rank of the $12\times 16$ matrix of the coefficients for this system is equal to 8, so that the set of Eqs.\ (7) determines 8 among the 16 probabilities $p1, p2,\ldots,p16$. So, for instance, we can get the following convenient solution of system (7) for which the set of variables $\mathcal{V}=\{p2,p3,p6,p7,p10,p11,p13,p16\}$ is given in terms of the remaining set of variables $\mathcal{U}=\{p1,p4,p5,p8,p9,p12,p14,p15\}$,
\begin{align}
p2 &= \frac{1}{2} (1-p1-p4+p5-p8-p9+p12+p14-p15),  \tag{8a}  \\
p3 &= \frac{1}{2} (1-p1-p4-p5+p8+p9-p12-p14+p15),  \tag{8b}  \\
p6 &= \frac{1}{2} (1+p1-p4-p5-p8-p9+p12+p14-p15),  \tag{8c}  \\
p7 &= \frac{1}{2} (1-p1+p4-p5-p8+p9-p12-p14+p15),  \tag{8d}  \\
p10 &= \frac{1}{2} (1-p1+p4+p5-p8-p9-p12+p14-p15),  \tag{8e}  \\
p11 &= \frac{1}{2} (1+p1-p4-p5+p8-p9-p12-p14+p15),  \tag{8f}  \\
p13 &= \frac{1}{2} (1-p1+p4+p5-p8+p9-p12-p14-p15),  \tag{8g}  \\
p16 &= \frac{1}{2} (1+p1-p4-p5+p8-p9+p12-p14-p15).  \tag{8h}  
\setcounter{equation}{8}
\end{align}
The basic relationships (8a)-(8h) between joint probabilities arise as a direct consequence of the fulfillment of the normalization condition and the causal communication constraint. It should be noted, however, that the conditions in Eq.\ (1) impose a lot of further constraints on their own. For example, the non-negativity of $p13$ in Eq.\ (8g) requires that
\begin{equation}
1+p4+p5+p9 \geq p1+p8+p12+p14+p15, 
\end{equation}
where, according to Eq.\ (1), all the probabilities $p\text{j}$ in the set $\mathcal{U}$ are assumed to fulfill the condition $0 \leq p\text{j} \leq 1$.$^{1}$ Similarly, other constraints like that of Eq.\ (9) arise when we demand that each of the probabilities $p\text{k}$ in the set $\mathcal{V}$ determined by Eqs.\ (8a)-(8h) fulfills the property $0\leq p\text{k} \leq 1$. Many more constraints can be established from Eqs.\ (8a)-(8h) by demanding the non-negativity of the sum of any combination of probabilities pertaining to the set $\mathcal{V}$. For example, the non-negativity of the sum $p2+p7$ requires that
\begin{equation}
p1+p8 \leq 1.
\end{equation}
Analogously, the non-negativity of the sum $p2+p3+p6+p7+p10+p11+p13+p16$ requires that
\begin{equation}
p1+p4+p5+p8+p9+p12+p14+p15 \leq 4,
\enlargethispage{2mm}
\end{equation}
and so on.

It is a well-established theoretical fact that the quantum mechanical probabilities satisfy the causal communication constraint. (A simple demonstration of the fact that the parameter independence conditions (3a)-(3b) are entailed by the quantum mechanical formalism is given in the Appendix, and more general demonstrations can be found, for example, in Ref.\ 1.$^{2}$) Therefore the quantum predictions for the probabilities $p1,p2,\ldots,p16$, \textit{whichever they might be}, should satisfy each of the constraints in Eqs.\ (8), as well as those induced by the non-negativity of single joint probabilities and the non-negativity of the sums of joint probabilities, such as the constraints in Eqs.\ (9)-(11). Quantum mechanics, however, is a rather peculiar statistical theory, and it places additional restrictions on the sets of probabilities it can produce, beyond those imposed by more general considerations like causality. Let us consider the CHSH sum of correlations, $\Delta = c(a_1,b_1) + c(a_1,b_2) + c(a_2,b_1) - c(a_2,b_2)$. Taking into account the normalization condition in Eq.\ (2), it is simple algebra to verify that this quantity can be equivalently expressed as
\begin{equation}
\Delta = 2(p1+p4+p5+p8+p9+p12+p14+p15-2).
\end{equation}
From the inequality (11), we can see that the maximum value for $\Delta$ is 4, and that this value will occur whenever $p1+p4+p5+p8+p9+p12+p14+p15 = 4$. Actually, there are two sets of values for the probabilities $p1, p2,\ldots,p16$ which provide the maximum absolute probabilistic limit for $\Delta_{\text{GP}}$, namely 4, and which, at the same time, satisfy the causality constraints in Eqs.\ (8a)-(8h). The first set of values is, $p\text{j}=1/2$ for all $p\text{j}\in \mathcal{U}$, and $p\text{k}=0$ for all $p\text{k}\in \mathcal{V}$; the second set is, $p\text{j}=0$ for all $p\text{j}\in \mathcal{U}$, and $p\text{k}=1/2$ for all $p\text{k}\in \mathcal{V}$. As we have said, the quantum mechanical probabilities obey both the inequality (11) and the constraints in Eqs.\ (8). However, such probabilities do not saturate the inequality (11). In fact, it can be shown that the maximum quantum prediction for the sum $p1+p4+p5+p8+p9+p12+p14+p15$ amounts to $2+\sqrt2$. There exist two sets of quantum mechanical values for the probabilities $p1, p2,\ldots,p16$ which provide the maximum absolute limit for $\Delta_{\text{QM}}$, namely $2\sqrt2$, and which, at the same time, are consistent with the conditions (8). The first set of values is, $p\text{j}=\frac{1}{8} (2+\sqrt2)$ for all $p\text{j}\in \mathcal{U}$, and $p\text{k}=\frac{1}{2} - \frac{1}{8} (2+\sqrt2)$ for all $p\text{k} \in \mathcal{V}$; the second set is, $p\text{j}=\frac{1}{2} - \frac{1}{8} (2+\sqrt2)$ for all $p\text{j}\in \mathcal{U}$, and $p\text{k}=\frac{1}{8} (2+\sqrt2)$ for all $p\text{k} \in \mathcal{V}$. Of course, as was already noted, the probabilistic limit $|\Delta_{\text{GP}}|=4$ entails that $c(a_1,b_1)= c(a_1,b_2)= c(a_2,b_1)= -c(a_2,b_2)= \pm 1$. In quantum mechanics the situation is radically different. In fact, it can be shown [21] that whenever we have the quantum predictions $c(a_1,b_1)= c(a_1,b_2)= c(a_2,b_1)= \pm 1$, then necessarily we must have $c(a_2,b_2)=\pm 1$ as well. In terms of joint probabilities this means that, for example, whenever quantum mechanics predicts that $p1=p4=p5=p8=p9=p12=1/2$, then necessarily it also predicts that $p14=p15=0$. It will be noted, incidentally, that this is the reason why it is not possible to construct a nonlocality argument of the Greenberger-Horne-Zeilinger type [22] for two spin-$\tfrac{1}{2}$ particles [21].\enlargethispage{3mm}

\section{Focusing on the case of Hardy's nonlocality theorem}

Let us now look at one of the constraints in Eqs.\ (8), for example, that in Eq.\ (8g). As, by assumption, the probabilities $p1,p8,p12,p14$, and $p15$ are non-negative, the following inequality must hold
\begin{equation}
2\,p13 -1 \leq p4+p5+p9 \, .
\end{equation}
This inequality is important for what follows because the four involved probabilities $p4, p5, p9,$ and $p13$ can be used to construct an argument for nonlocality of the type invented by Hardy [15].$^{3}$ Actually, inequality (13) was already derived by Mermin\linebreak in Ref.\ 8 (see, in particular, Eq.\ (9) and Appendix A of [8]). The derivation presented in this paper, however, has been made within a more general framework. Indeed, the inequality (13) arises here as an immediate by-product of the basic constraint, Eq.\ (8g), $2\,p13 -1 =p4+p5+p9-p1-p8-p12-p14-p15$. In this respect, from a pedagogical point of view, our derivation has the advantage of being more straightforward. Moreover, thanks to this general framework (the ``CHSH framework''), we have been able to quickly identify (see note 3) the eight sets of probabilities which can lead to a Hardy-type nonlocality contradiction, along with the constraint of the type (13) that causality imposes on each of these sets. In what follows we pick out the set of probabilities appearing in Eq.\ (13), $\{p4,p5,p9,p13\}$, although, naturally, any one of the remaining sets could be equally well considered.$^{4}$

Hardy's nonlocality argument applies to the case in which the three probabilities $p4$, $p5$, and $p9$ vanish while the probability $p13$ does not. For this case the inequality (13) reduces to $p13 \leq 1/2$. Thus, the non-negativity condition, $0\leq p13$, combined with the causality constraint, $p13 \leq 1/2$, determine the range of values
\begin{equation}
0 \leq p13 \leq 1/2 \, ,
\end{equation}
within which the probability $p13$ can vary without violating the causal communication constraint for the case that $p4=p5=p9=0$. It is easy to calculate the absolute value taken by the quantity $\Delta$ in the probabilistic limiting case in which $p13 =1/2$ and $p4=p5=p9=0$. Indeed, it is clear from Eq.\ (8g) that, whenever we have $p13=1/2$ and $p4=p5=p9=0$, then necessarily $p1=p8=p12=p14=p15=0$. Hence, since $p\text{j}=0$ for all $p\text{j}\in \mathcal{U}$, Eq.\ (12) tells us that $|\Delta_{\text{GP}}|=4$. Of course, from Eqs.\ (8a)-(8h), we also have that $p\text{k}=1/2$ for all $p\text{k} \in \mathcal{V}$. The situation for quantum mechanics is, once again, quite different. Although quantum mechanics does satisfy the constraint in Eq.\ (14) for the case in which $p4=p5=p9=0$, the quantum prediction for $p13$ does not saturate the upper bound of the inequality (14). Indeed, it can be shown [8,15,16] that the maximum attainable value for $p13$ predicted by quantum mechanics when $p4=p5=p9=0$ is $\tau^{-5}$, with $\tau$ being the golden mean, $\frac{1}{2}(1+\sqrt{5})$. This gives a maximum value of $p13=0.090\,\, 17$.

At this point we derive an immediate but important relationship between the absolute value of the quantity $\Delta$ and the probability $p13$, which applies in the considered case in which $p4=p5=p9=0$. So, from Eq.\ (12), we have that, for this case, $\Delta = 2(p1+p8+p12+p14+p15-2)$. On the other hand, when we put $p4=p5=p9=0$ in Eq.\ (8g), we deduce that $p1+p8+p12+p14+p15=1-2\, p13$. Hence we obtain
\begin{equation}
|\Delta|= 2 + 4\, p13 \, .
\end{equation}
It should be stressed that the identity (15) arises as a direct consequence of the causality constraint in Eq.\ (8g). Therefore, relation (15) should be fulfilled by the quantum mechanical probabilities if these are to satisfy the causal communication constraint, Eqs.\ (3). It can be shown [21] that, in fact, the quantum theoretic predictions do satisfy the identity (15) for the case that the probabilities $p4$, $p5$, and $p9$ are made to vanish. Furthermore, it is worth noting that the relation (15) gives a new insight into the rationale of the causality constraint in Eq.\ (14). Indeed, if $p13$ were allowed to be greater than 1/2 when $p4=p5=p9=0$, then the CHSH sum of correlations on the left side of (15) would be greater than 4, which is impossible by the very definition of $\Delta$. A straightforward implication of Eq.\ (15) is that $|\Delta|>2$ whenever $p4=p5=p9=0$ and $p13>0$, and that $|\Delta|=2$ whenever $p4= p5=p9=p13=0$. On the other hand, from the results expounded in this and in the preceding paragraph, it is evident that relativistic causality does not by itself restrict the maximum CHSH sum of quantum correlations to $2 + 4\tau^{-5}$ for the case in which $p4=p5=p9=0$, since, as we have seen, we can indeed have a general probabilistic situation in which $p4=p5=p9=0$ and $|\Delta_{\text{GP}}|=4$. Of course, from Eq.\ (15), this will happen whenever $p13=1/2$. We also note that, for the special case in which the state of the composite system is described by a maximally entangled state,$^{5}$ quantum mechanics predicts that $p13=0$ whenever we have $p4=p5=p9=0$. Thus the maximally entangled state cannot be used to exhibit Hardy-type nonlocality, in spite of the fact that this class of states yields the maximum quantum violation, $\left| \Delta_{\text{QM}} \right| = 2\sqrt2$, of the CHSH inequality [23].  

The quantum mechanical restriction (see Eq.\ (15)), $|\Delta_{\text{QM}}| \leq 2+4\tau^{-5}=2.360\,\, 68$, which applies to the case where $p4=p5=p9=0$, is more stringent than the restriction $|\Delta_{\text{QM}}| \leq 2\sqrt{2}=2.828\,\,42$ obtained when no a priori conditions (other than those in Eqs.\ (1)-(3)) are imposed on the probabilities $p1, p2,\ldots,p16$. This is the reason why Hardy's theorem does not lead to a more definitive experimental test of quantum nonlocality than the ones already performed. However, it should be appreciated that, for the case in which $p4=p5=p9=0$, quantum mechanics predicts an amount of violation of the CHSH inequality, $|\Delta_{\text{LHV}}|\leq 2$, which is comparatively larger than that predicted for the CH type inequality [8], $p13 \leq p4+p5+p9$. Indeed, for the considered case, and in accordance with quantum mechanics, the quantity $\Delta$ can reach a value as large as $|\Delta_{\text{QM}}|=2.360\,\,68$, so that the maximum violation of the CHSH inequality predicted by quantum mechanics is $2.360\,\,68 \leq 2$. Normalizing this latter inequality to 2, we obtain $(2.360\,\,68 -2)/2 \leq0$, or $0.180\,\,34 \leq0$.$^{6}$ On the other hand, the corresponding quantum violation of the CH inequality is $0.090\,\,17 \leq 0$. It is therefore concluded that experiments based on the CHSH inequality should give a more conclusive, clear-cut experimental verification of Hardy's nonlocality than the ones based on the CH inequality, \textit{provided} that the magnitude of the experimental error is the same for both kinds of experiments.

We conclude by briefly examining the middle ground, \mbox{$2.360\,\,68 \leq \Delta \leq 4$}, between the upper limits imposed by quantum mechanics and relativistic causality within the context of Hardy's nonlocality theorem. To this end, we write down the causality constraint in Eq.\ (8g) that obtains for the case in which $p4=p5=p9=0$,
\begin{equation}
p1+p8+p12+p14+p15 =1-2\, p13 \, . 
\end{equation}
Since quantum mechanics requires that $0\leq p13 \leq 0.090\,\,17$ when $p4=p5=p9=0$, then the quantum prediction for the sum of probabilities on the left-hand side of (16) is constrained to obey the inequality
\begin{equation}
0.819\,\,66 \leq \Sigma_{\text{QM}} \leq  1 \, , 
\end{equation}
where the symbol $\Sigma$ stands for the sum $p1+p8+p12+p14+p15$. This inequality translates into the following one, $2\leq |\Delta_{\text{QM}}| \leq 2.360\,\,68$, with $|\Delta_{\text{QM}}|$ reaching its maximum value $2.360\,\,68$ whenever the sum $\Sigma$ equals $0.819\,\,66$. The quantum mechanical probabilities giving this maximum value are $p1=\tau^{-3}$ and $p8=p12=p14=p15=\tau^{-4}$. On the other hand, for a general probabilistic theory, and for the case that $p4=p5=p9=0$, the probability $p13$ is required to satisfy the less stringent condition $0\leq p13 \leq 1/2$, and thus we now have the inequality
\begin{equation}
0 \leq \Sigma_{\text{GP}} \leq  1 \, ,
\end{equation}
which translates into the inequality $2\leq |\Delta_{\text{GP}}| \leq 4$. The transition from the quantum domain to the general probabilistic domain happens when the value of $\Sigma$ drops below the threshold $0.819\,\,66$. Thus the lower bound of the inequality (17) provides a tool to discriminate between quantum mechanics and general probabilistic theories. Indeed, if the observable quantity $\Sigma$ were found experimentally to lie within the quantum mechanically forbidden interval, \mbox{$0\leq \Sigma < 0.819\,\,66$}, for a situation in which the probabilities $p4$, $p5$, and $p9$ have been made to vanish, then quantum mechanics would prove wrong. Of course we do not expect any violation of quantum mechanics to be found. It is to be emphasized, however, that a violation of the lower bound of the inequality (17) for the case where $p4=p5=p9=0$, is not at all forbidden by the general requirement of relativistic causality. A survey of what is known about correlations in physical systems, along with additional restrictions on the kinds of correlations which are allowed by quantum mechanics, can be found in \mbox{Ref.\ 4}.

\section{Conclusion and summary}

A final remark is in order about the general nature of the extra constraints added by quantum mechanics, as compared with those entailed by general probabilistic theories. From the preceding  paragraphs it should be clear that, for the considered CHSH and CH tests of nonlocality, the additional constraints implied by quantum mechanics essentially entail a weakening of the correlations it can produce. So, for example, as we have seen, the quantum mechanical probabilities do not saturate the inequality in Eq.\ (11), so that the quantum prediction of the CHSH sum of correlations, Eq.\ (12), is bounded in absolute value by $2\sqrt{2}$ (Cirel'son limit), instead of $4$. Moreover, this weakening is not at all determined by the requirement of relativistic causality since it is theoretically possible to have (hypothetical) stronger-than-quantum correlations preserving relativistic causality. It should be noticed, however, that, generally speaking, the constraints added by quantum mechanics do not prevent a given joint probability from attaining a value of unity. Indeed, if the system is prepared in a suitable product state, we can always make a given probability, say, $p1$, to be equal to unity, yet having the constraint $|\Delta_{\text{QM}}| \leq 2\sqrt{2}$. (In fact, for product states, we have the more stringent bound $|\Delta_{\text{QM}}| \leq 2$.)

In summary, in the present paper we have determined a number of basic restrictions (cf.\ Eqs.\ (8)) on the joint probabilities involved in an experiment of the Einstein-Podolsky-Rosen-Bohm type, which develop when these are required to satisfy the normalization condition and the causal communication constraint. These restrictions are, therefore, rather general and should be fulfilled by any physical theory consistent with relativistic causality. Further constraints arise when the sums of joint probabilities are required to satisfy the non-negativity condition. We have also considered the conceptually important case in which three specific probabilities are set to zero. This allows a formulation of Hardy's nonlocality theorem within the framework of the CHSH inequality. For the abovementioned case, we have obtained the relevant constraints imposed by a general probabilistic theory, on the one hand, and by quantum mechanics, on the other hand, and we have derived a simple inequality discriminating between the two kinds of theories.

\vspace{.2cm}
\textbf{Acknowledgments} --- The author wishes to thank the anonymous referees for their valuable comments which led to an improvement of an earlier version of this paper.

\newpage
\centerline{\textbf{{\Large Appendix}}}
\vspace{.5cm}
To show how the parameter independence condition embodied in the second equality of Eq.\ (3a) arises as a consequence of the formalism of quantum mechanics, we introduce the quantum mechanical operators $\hat{a}_1$, $\hat{a}_2$, $\hat{b}_1$, and $\hat{b}_2$, which are defined through the eigenvalue equations
\begin{align}
\hat{a}_j |m;\hat{a}_j \rangle &=\, m \,|m;\hat{a}_j \rangle,  \nonumber  \\
\hat{b}_k |n;\hat{b}_k \rangle &=\, n \,|n;\hat{b}_k \rangle,  \tag{A1}
\end{align}
where $j,k = 1 \text{ or } 2$, and $m,n=\pm 1$. The eigenvectors of $\hat{a}_j$ satisfy the orthogonality relation $\langle +;\hat{a}_j | -;\hat{a}_j \rangle =0$ and, furthermore, they are normalized to length 1, that is, $\langle +;\hat{a}_j | +;\hat{a}_j \rangle = \langle -;\hat{a}_j | -;\hat{a}_j \rangle =1$, with similar relations holding for the eigenvectors of $\hat{b}_k$. The operators $\hat{a}_j$ and $\hat{b}_k$ correspond to the measurement of the variables $a_j$ and $b_k$, respectively. Then, in the simplest (idealized) situation in which the quantum state describing the composite system is the pure state $|\psi \rangle$, the probability of getting the outcomes $a_j =m$ and $b_k =n$ in a joint measurement of $\hat{a}_j$ and $\hat{b}_k$ is
\begin{equation}
p(a_j=m, b_k=n) = \langle\psi|\, \hat{p}_m (\hat{a}_j)\, \hat{p}_n (\hat{b}_k)\, |\psi\rangle ,
\tag{A2}
\end{equation}
where $\hat{p}_m (\hat{a}_j)$ and $\hat{p}_n (\hat{b}_k)$ are the projection operators $\hat{p}_m (\hat{a}_j) = |m;\hat{a}_j\rangle\langle m;\hat{a}_j|$ and $\hat{p}_n (\hat{b}_k) = |n;\hat{b}_k\rangle\langle n;\hat{b}_k|$. We note that the operators $\hat{p}_m (\hat{a}_j)$ and $\hat{p}_n (\hat{b}_k)$ are assumed to be compatible (that is, mutually commuting) in the case that the space-time events corresponding to the measurement of $\hat{a}_j$ and $\hat{b}_k$ are spacelike separated, so that, for such a case, the product $\hat{p}_m (\hat{a}_j)\,\hat{p}_n (\hat{b}_k)$ is a well-defined observable operator. We thus have
\begin{align}
p(a_j=m) &= \sum_{n=\pm1} \langle\psi| \,\hat{p}_m (\hat{a}_j)\, \hat{p}_n (\hat{b}_k) \, |\psi\rangle  \nonumber \\
&= \langle\psi| \,\hat{p}_m (\hat{a}_j) \left(\sum_{n=\pm1}\hat{p}_n (\hat{b}_k)\right) |\psi\rangle .   \tag{A3}
\end{align}
Now, since the sum of projectors $\hat{p}_{+} (\hat{b}_k)+\hat{p}_{-} (\hat{b}_k)$ is the identity operator acting on the two-dimensional Hilbert space pertaining to part $B$, the expression in Eq.\ (A3) reduces to
\begin{equation}
p(a_j=m)= \langle\psi|\, \hat{p}_m (\hat{a}_j)\,|\psi\rangle ,
\tag{A4}
\end{equation}
which, clearly, is independent of the choice of measurement ($b_1$ or $b_2$) performed on part $B$. Hence relation (3a) follows. In fact, regarding the value of the quantum mechanical probability $p(a_j=m)$, it is immaterial whether some measurement is actually performed on $B$ or not. An entirely analogous argument could be established to show the independence of the quantum mechanical probability $p(b_k=n)$ from any measurement parameter concerning part $A$ of the system.

\vspace{.5cm}
\center

\newpage
\centerline{\textbf{{\Large Notes}}}
\begin{enumerate}
\item For the sake of completeness, it should be mentioned that various authors have entertained the logical possibility of solving the EPR paradox by considering extended probability measures (including negative ones). See, for example, the papers quoted in Ref.\ 19.

\item From the experimental side, there has been a recent test of the CHSH inequality conducted by Weihs \textit{et al.} (see the first paper in Ref.\ 20) which, to date,\linebreak is the only one experiment to force a violation of Bell's inequality under truly spacelike separation of the individual measurement processes of the two involved observers. This experiment can be considered as an update with present-day technology of the classic third experiment by Aspect and co-workers (see the second paper in [20]). To date, however, all performed experiments testing Bell inequalities (including that of Weihs \textit{et al.}) rely on one or another sort of auxiliary assumption (like the fair-sampling assumption) in order to deal with the detection efficiency loophole. Further experimental work testing quantum correlations in relativistic configurations is reported in the third paper of [20].

\item It is to be noted that Hardy's nonlocality argument can be constructed out of other suitable sets of four probabilities, for example, out of the set $\{p1,p8,\linebreak p12,p16\}$. The relevant constraint imposed by relativistic causality on this set will be (see Eq.\ (8h)), $2\,p16-1 \leq p1+p8+p12$. Indeed, there exists a total of eight sets of four probabilities for which it is possible to develop Hardy's nonlocality theorem. Each of these sets is associated with one specific relation in Eqs.\ (8a)-(8h). For any given relation, the four probabilities forming the set are the three on the right-hand side with positive sign, plus the one on the left-hand side. So, for example, the set associated with relation (8a) is $\{p2,p5,p12,p14\}$, the set associated with relation (8b) is $\{p3,p8,p9,p15\}$, etc.

\item For ease of comparison with the work of Mermin in Ref.\ 8, we quote here the translation between Mermin's notation and ours for the probabilities appearing in Eq.\ (13): $p(11RR) \equiv p4$, $p(12GG) \equiv p5$, $p(21GG) \equiv p9$, and $p(22GG) \equiv p13$.

\item The paradigm of maximally entangled state for a system composed of two subsystems is the singlet state of two spin-$\frac{1}{2}$ particles, $|\psi \rangle =(1/\sqrt{2})(|\uparrow\rangle_{1} |\downarrow\rangle_{2}\linebreak -  |\downarrow\rangle_{1} |\uparrow\rangle_{2})$, or its photon analog (we omit the spatial wave function of the state). $|\uparrow\rangle_{1}$, $|\downarrow\rangle_{1}$, $|\uparrow\rangle_{2}$, and $|\downarrow\rangle_{2}$ represent the spin states of the two particles (polarized along a common $z$-axis). The maximally entangled state is a symmetric state in that the two coefficients of the (biorthogonal) superposition have the same modulus. For the maximally entangled state, the quantum mechanical probabilities satisfy the symmetry relations, $p1=p4$, $p2=p3$, $p5=p8$, $p6=p7$, $p9=p12$, $p10=p11$, $p13=p16$, and $p14=p15$. In the case of two spin-$\frac{1}{2}$ particles, the variables $a_1$, $a_2$, $b_1$, and $b_2$ denote spin measurements (with outcomes $+1$ or $-1$) along different directions for each of the particles.

\item Another way of obtaining this quantum violation is the following. Assuming without loss of generality that $p1+p4+p5+p8+p9+p12+p14+p15 \leq 2$, the CHSH inequality, $|\Delta_{\text{LHV}}|\leq 2$, reads as (see Eq.\ (12)), $2-p1-p4-p5-p8-p9-p12-p14-p15 \leq 1$ or, equivalently, $1-p1-p8-p12-p14-p15 \leq p4+p5+p9$. It is this latter inequality which can be compared with the CH type inequality, $p13 \leq p4+p5+p9$. When $p4=p5=p9=0$, the above inequality reduces to $1-p1-p8-p12-p14-p15 \leq 0$. Now, from Eq.\ (16) (see below), we finally obtain $2\, p13 \leq 0$. The maximum quantum violation of this inequality is then $0.180\,\,34 \leq 0$. (Please note that, for a theory of local hidden variables, we have that $p13=0$ whenever $p4=p5=p9=0$, and thus the inequality $2 \, p13 \leq 0$ is satisfied.)

\end{enumerate}

\end{document}